\documentclass[twocolumn,showpacs,aps]{revtex4}
\usepackage{graphicx}

\begin{document}

\title{Oscillatory wave fronts in chains of coupled nonlinear oscillators}
\author{A. Carpio\cite{carpio:email} }
\affiliation{Departamento de Matem\'{a}tica Aplicada, Universidad Complutense de
Madrid, 28040 Madrid, Spain}
\author{ L. L. Bonilla\cite{bonilla:email}}
\affiliation{Departamento de Matem\'aticas, Universidad Carlos III de Madrid, Avda.\
Universidad 30, E-28911 Legan{\'e}s, Spain}
\date{ \today  }

\begin{abstract} 
Wave front pinning and propagation in damped chains of coupled oscillators are
studied. There are two important thresholds for an applied constant stress $F$: for
$|F|<F_{cd}$ (dynamic Peierls stress), wave fronts fail to propagate, for $F_{cd} <
|F| < F_{cs}$ stable static and moving wave fronts coexist, and for $|F| > F_{cs}$
(static Peierls stress) there are only stable moving wave fronts. For piecewise
linear models, extending an exact method of Atkinson and Cabrera's to chains with
damped dynamics corroborates this description. For smooth nonlinearities, an
approximate analytical description is found by means of the active point theory.
Generically for small or zero damping, stable wave front profiles are non-monotone and
become wavy (oscillatory) in one of their tails.
\end{abstract}

\pacs{45.05.+x; 05.45.-a; 83.60.Uv }
\maketitle

\section{Introduction}
Wave fronts and pulses play important roles in many physical systems. Examples abound: the
motion of dislocations \cite{fk,nab67} or cracks \cite{sle81} in crystalline
materials, atoms adsorbed on a periodic substrate \cite{cha95}, the motion of
electric field domains and domain walls in semiconductor superlattices
\cite{cba01,bon02}, pulse propagation through myelinated nerves \cite{sleeman},
pulse propagation through cardiac cells \cite{kee98}, etc. Furthermore, these
localized waves often play an important role in Statistical Mechanics \cite{nel02} or
Quantum Field Theory \cite{rebbi}. When wave fronts or pulses are solutions of
spatially discrete systems, they often fail to propagate unless an external force or
parameter surpasses a critical value \cite{cb01}. Wave front pinning in discrete
systems may be related to such different physical phenomena as the existence of
Peierls stresses in continuum mechanics \cite{hob65} or the relocation of electric
field domains in semiconductor superlattices \cite{bon02}. In the continuum limit,
the width of the pinning interval (range of the external force for which wave fronts
fail to propagate) tends to zero exponentially fast and many authors have calculated
the critical force for different models in this limit \cite{ind58,cah60,nab67,kin01}.

Not surprisingly, wave front motion and pinning are different depending on the
dynamics describing the model at hand. To be precise, let us consider a chain of
nonlinear oscillators, diffusively coupled and subject to an external force $F$ which
acts as a control parameter:
\begin{eqnarray}
m {d^2 u_n\over d\tau^2} + {du_{n}\over d\tau} = u_{n+1} -2 u_n + u_{n-1} -A\, g(u_n)
+ F.
\label{amort} 
\end{eqnarray} 
Typical nonlinearities $g(u)=V'(u)$ are cubic, such that $A\, g(u)-F$ has  three
zeros, $U_1(F/A) <U_2(F/A) <U_3(F/A)$ in a certain force interval ($g'(U_i(F/A)) >0$
for $i=1,3$, $g'(U_2(F/A))<0$). Moreover, $g(u)$ is symmetric with respect to
$U_2(0)$. Examples are the overdamped Frenkel-Kontorova (FK) model ($g=\sin u$)
\cite{fk} and the quartic double well potential ($V=(u^2-1)^2 /4$) \cite{cah60}. $A>0$
measures the strength of the coupling and $m$ the relative strength of inertial and
friction terms. Wave front solutions $u_n=w(n-c\tau)$ join the two stable constant
states $U_1(F/A)$ and $U_3(F/A)$ (or viceversa) as $n$ increases from $-\infty$ to
$\infty$. 

Consider the extreme cases of conservative ($m=\infty$) and overdamped dynamics
($m=0$). In the overdamped case, wave fronts generically either move if $|F|>F_c>0$
or are pinned if $|F|\leq F_c$ \cite{cb01}. The depinning transition at $F_c$ was
described by Carpio and Bonilla \cite{cb01} for large and moderate values of $A$, by
King and Chapman \cite{kin01} in the continuum limit $A\to 0$, and by F\'ath
\cite{fat98} for a piecewise linear $g(u)$. In the conservative case and for generic
cubic nonlinearities $g(u)$, there are two critical forces $F_{cd}$ and $F_{cs}$ with
$0<F_{cd} <F_{cs} =F_c$. Wave fronts may propagate stably for $|F|> F_{cd}$ but there
are stable stationary (pinned) wave fronts if $|F|<F_c$. Thus pinned and moving wave
fronts may coexist if $F_{cd}<|F| < F_{cs}$. The values $F_{cs}$ and $F_{cd}$
correspond to the static and dynamic Peierls stresses of the literature on
dislocations \cite{nab67}. Atkinson and Cabrera found exact expressions for the wave
fronts corresponding to a piecewise linear $g(u)$ and calculated the relationship
between $F$ and wave front velocity \cite{atk65}. An approximate theory was found
somewhat earlier by Weiner \cite{wei64}. More recently, Schmidt \cite{sch79} and later
authors \cite{bre97,fla99} found exact monotone wave fronts of conservative or
overdamped systems by constructing models with nonlinearities such that the desired
wave fronts were solutions of the models. In particular, Flach et al \cite{fla99}
showed coexistence between moving and pinned monotone wave fronts of a discrete
system with conservative dynamics and $F=0$. On the other hand, for a sine
nonlinearity and $F=0$, the numerical computations of Peyrard and Kruskal
\cite{pey84} show that an initial profile close to the continuum sine Gordon soliton
loses energy via emission of phonons and it becomes pinned after a sufficiently long
time interval. They also found stable moving wave fronts for small positive $F$,
consistent with our previous statement that, generically, $F_{cd}>0$. 

In this paper, we study the wave fronts of the damped system (\ref{amort}) and the 
transitions at $F_{cd}$ and $F_{cs}$. In contrast with previous work, we find wave
fronts that are non-monotone, presenting wavy tails at one or both sides of a
transition region in which the profile $w(n-c\tau)$ jumps an amount close to
$[U_3(F/A) - U_1(F/A)]$. We call them {\em wavy wave fronts}. These fronts with wavy
profiles persist even in the conservative limit ($m\to\infty$) \cite{definition}, and
in fact Atkinson and Cabrera's wave fronts are also wavy, as these authors would have
found out had they depicted their exact expression graphically. In the overdamped
limit $m\to 0$, $F_{cd} \to F_c$ and the wave front profiles become monotone. We have
thus arrived to a general picture of wave fronts in discrete chains of coupled
nonlinear oscillators with $m>0$. 

The rest of the paper is organized as follows. Section \ref{sec:cabrera} considers
Eq.\ (\ref{amort}) with a piecewise linear $g(u)$. We find exact formulas for the
wave front profiles in the general damped case following the method of Atkinson and
Cabrera's \cite{atk65}. These profiles are often wavy and they are asymptotically
stable in the damped case. It is important to obtain them for two reasons: (i) there
are very few exact wave front solutions that are non-monotone, and (ii) in the limit
of large inertia, it is hard to discriminate numerically between wavy wave fronts
traveling with different velocities or having different profiles. Exact solutions make
good benchmarks for numerical methods. The results for the damped model with a
generic cubic nonlinearity are presented in Section \ref{sec:generic}. We calculate
the static and dynamic Peierls stresses for typical values of $A$ and $m$. A
characterization of these stresses is given in terms of our active point theory.
Sec.~\ref{sec:conclusions} contains a discussion of our results. 

\section{Explicit construction of wave front profiles}
\label{sec:cabrera}

\begin{figure}
\begin{center}
\includegraphics[width=8cm]{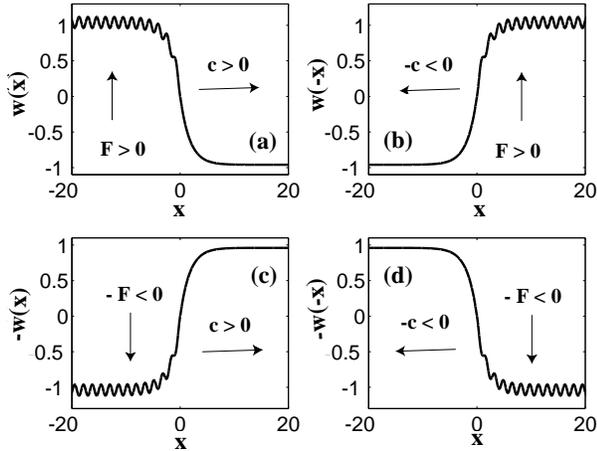} 
\end{center}
\caption{Symmetries in the wavefront solutions for $A=0.25$, $\alpha=0$, $c= 0.5.$ and
$F=0.009$. }
\label{figura1}
\end{figure}

Let us rescale time in Eq.\ (\ref{amort}), so that $t= \tau/\sqrt{m}$, and consider
a piecewise linear $g(u)$:
\begin{eqnarray} 
{d^2 u_n\over dt^2} + \alpha {du_n\over dt} = u_{n+1}-2 u_n + u_{n-1} - A g(u_n)
+F, \label{cab1} \\ 
g(u_n)= \left\{ \begin{array}{l} u_n + 1, \quad\mbox{for} \;  u_n<0 ,
\\ u_n - 1,\quad\mbox{for} \; u_n\geq 0, \end{array} \right. 
\label{cab0}
\end{eqnarray} 
where $\alpha=1/\sqrt{m}$. Notice that $g(u)= u+1 - 2\, H(u)$, where $H(x)=1$ for
$x>0$ and $H(x)=0$ for $x<0$ is the Heaviside unit step function. Let us consider a
smooth wave front profile $u_n = w(x)\equiv v(x) - 1$, $x=n-ct$, moving rigidly
with velocity $c$. We center the wave front so as to have $w(0)=0$. Taking into
account  that $g(u)$ is an odd function and using the front profile $u_n(t)=
w(n-ct)$, we can see that the following transformations leave Eq.\ (\ref{cab1})
for $w(x)$ invariant: 
\begin{eqnarray} 
&(x,w,c,F) \to (-x,w,-c,F) \nonumber\\
&(x,w,c,F) \to (x,-w,c,-F) \nonumber\\
&(x,w,c,F) \to (-x,-w,-c,-F)  .  \label{cab3}
\end{eqnarray} 
Let us consider now the case $F>0$, $c>0$ and $w'(0)<0$, i.e., a wave front profile
that decreases in the transition region about $x=0$. The transformations (\ref{cab3})
yield a profile with: (i) $c<0$, $F>0$ increasing in the transition region, (ii)
$c>0$, $F<0$ increasing in the transition region, and (iii) $c<0$, $F<0$ decreasing
in the transition region. Thus we find that sgn$w=-$sgn$(xcF)$, $g(w)=w+1 - 2\, H(- x
$sgn$(cF))$, and we can restrict ourselves to considering the case $F>0$, $c>0$ and
$w'(0)<0$: All other three possible cases can be obtained from our results by using
Eq.\ (\ref{cab3}); see Figure \ref{figura1}. 

\begin{figure}
\begin{center}
\includegraphics[width=8cm]{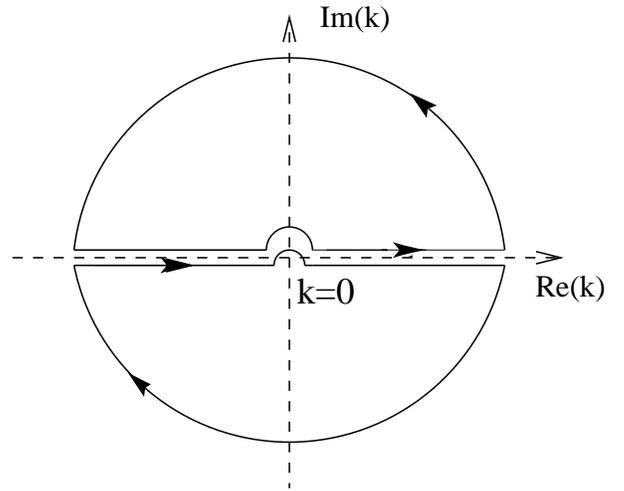} 
\end{center}
\caption{Contour for the Heaviside step function (\ref{cab4}) and the integral
formula (\ref{cab5})-(\ref{cab6}) when $\alpha\neq 0$.}
\label{figura2} 
\end{figure}

The wave front profile $v(x)= w(x)+1$ satisfies:
\begin{eqnarray} 
c^2 v''(x)- \alpha c v'(x) - [v(x+1)-2v(x)+v(x-1)]  \nonumber\\
+A\, v(x) =  2A\, H(-\mbox{sgn}(cF)\, x ) + F,  \label{cab2}
\end{eqnarray} 
with $v(0)=1$. We can calculate $v(x)$ by using the contour integral expression for
the step function:
\begin{eqnarray} 
H(-x)= -{1\over 2 \pi i} \int_{C} {e^{ik\, x}\over k}\, dk. \label{cab4}
\end{eqnarray}
Here $C$ runs over the real axis in the complex $k$ plane passing above the pole at
$k=0$ as in Fig.~\ref{figura2}. For $x>0$ (resp.\ $x<0$), $C$ is closed by a
semicircle in the upper (resp.\ lower) half plane oriented counterclockwise (resp.\
clockwise). 

\begin{figure}
\begin{center}
\includegraphics[width=8cm]{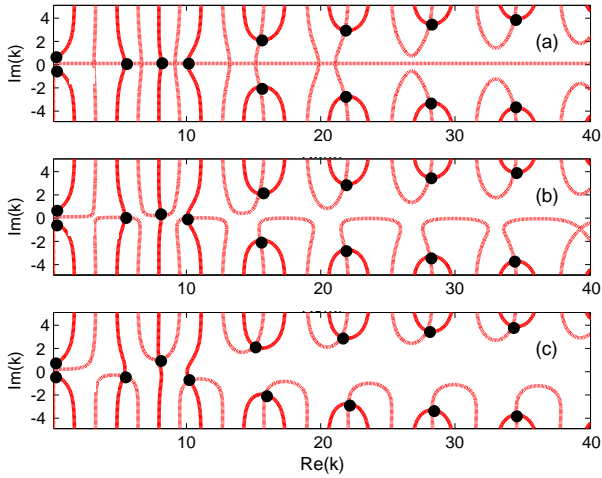} 
\end{center}
\caption{Complex poles as intersection of the curves Re$L(k,\alpha)=0$ and Im$L(k,
\alpha) =0$ when $c=0.2$, $A=0.25$ and: (a) $\alpha=0$, (b) $\alpha=0.1$, (c)
$\alpha=1$.}
\label{figura4}
\end{figure}

\begin{figure}
\begin{center}
\includegraphics[width=8cm]{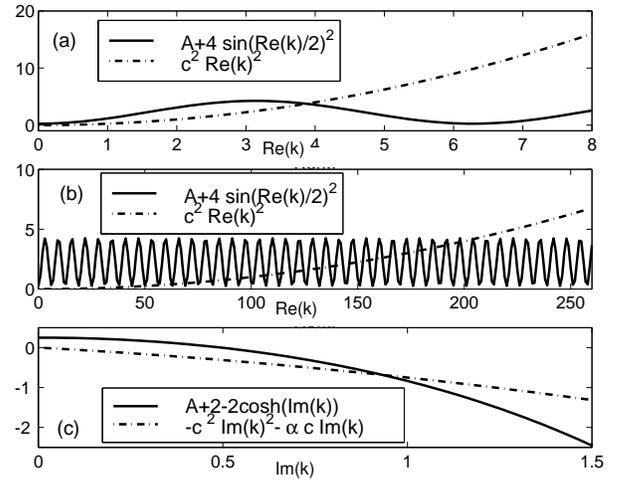}
\end{center}
\caption{Real poles in the conservative case $\alpha=0$ with $A=0.25$: (a) $c=0.5$
(b) $c=0.01$; Purely imaginary poles when $A=0.25$ (c) $c=0.5$, $\alpha=1$}
\label{figura3a}
\end{figure}

\begin{figure}
\begin{center}
\includegraphics[width=8cm]{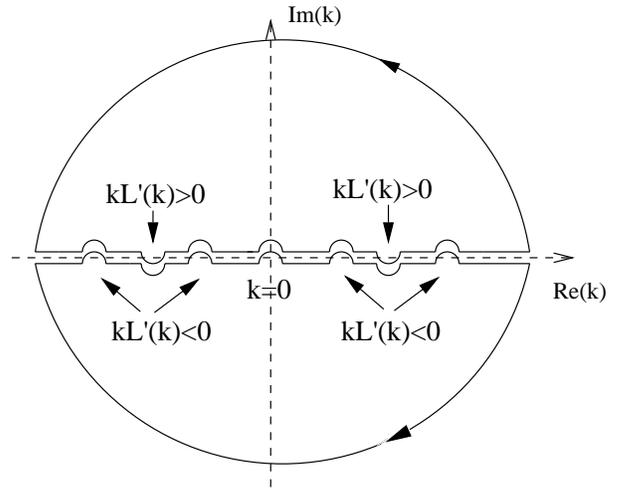} 
\end{center}
\caption{Contour for the integral formula (\ref{cab5})-(\ref{cab6}) in the
conservative case $\alpha=0$ for $c>0$ and $F>0$.}
\label{figura3}
\end{figure}

Then the solution of Eq.\ (\ref{cab2}) is
\begin{eqnarray} 
v(x)= {F\over A} - {A\over \pi i} \int_{C} {\exp[ik\; \mbox{sgn}(cF) x]\, dk\over k\,
L(k,\alpha)} , \label{cab5}\\ 
L(k,\alpha) = A+4 \sin^2\left({k\over 2}\right) -k^2 c^2 - ik |c| \alpha\,
\mbox{sgn}(F).
\label{cab6}
\end{eqnarray} 
All the zeros of the function $L(k,\alpha)$ given by (\ref{cab5}) are complex for
$\alpha>0$, and they correspond to exponentially localized modes. The nonzero poles
of the integrand in Eq.\ (\ref{cab5}) can be found graphically by plotting the curves
Re$L(k,\alpha)=0$ and Im$L(k,\alpha)=0$ in the complex $k$-plane, as depicted in Fig
\ref{figura4}. When $\alpha\to 0$, a finite number of poles tend to the real axis,
whereas infinitely many keep a nonzero imaginary part even at $\alpha=0$. The poles
on the real axis correspond to radiation modes, cause oscillations in the wave front
tails, and their number increases as $c$ decreases; see Figs. \ref{figura3a}(a) and
(b). The purely imaginary poles of Figures \ref{figura4} and \ref{figura3a}(c) yield
the central monotone part of the wave front profiles. For $\alpha=0$, the integration
contour in Eq.\ (\ref{cab5}) avoids poles on the real axis according to a criterion
due to Atkinson and Cabrera \cite{atk65}, shown in Figure \ref{figura3} and derived
later on this Section. We will use (\ref{cab5}) and the method of residues to
construct profiles satisfying $v(x)>1$ for $x<0$ and $v(x)<1$ for $x>0$. Notice that
we can obtain a complex dispersion relation between $\omega = kc$ and $k$ from
$L(k,\alpha)=0$. The contour choice and the fact that $\alpha>0$ give rise to an
exponential decay of $v(x)$ to its asigned values at $x=\pm \infty$. When $\alpha=0$,
the wave fronts may exhibit undamped oscillations extending all the way to infinity.

The condition $v(0)=1$ yields a relationship between the wave front velocity $c$ and
the external force $F$:
\begin{eqnarray} 
1 &=& {F\over A} - {A\over \pi i} \int_{C} {dk \over k  L(k,\alpha)} \nonumber\\
&=& {F\over A} - 
\sum_{L(p,\alpha)=0, Im(p)>0} {2A\over p\, L_k(p,\alpha)}, \label{cab7}
\end{eqnarray} 
where we have assumed that $cF>0$ and $v'(1)<0$. The resulting function $F(c)$ can be
calculated by computing this series of residues numerically. Once $F(c)$ is known,
Eq.\ (\ref{cab5}) can be used to compute the wave front profiles for a pair
$(c,F(c))$. We shall now show how this construction works out for $\alpha=0$,
$\alpha=\infty$ and for $\alpha$ finite.

\subsection{Conservative case: $\alpha=0$}
It is instructive to see what happens in the conservative limit $\alpha\to 0+$. In 
this case, $L(k,0)=0$ has real solutions, and we need a criterion to move the
contour $C$ above or below the corresponding poles in the integral (\ref{cab5}). To
obtain it, we shall use the notation $L(k)=L(k,0)$. Let $k_0$ be a real zero of
$L(k)$. The complex zero of $L(k,\alpha)$ that becomes $k_0$ at $\alpha=0$ satisfies
$0 = L_k(k_0,0)\, (k-k_0) + L_\alpha(k_0, 0)\,\alpha+ \ldots$, which yields $k\sim
k_0 - L_\alpha(k_0,0) \alpha/ L_k(k_0,0)$, that is,
\begin{eqnarray} 
k \sim k_0 + {i\alpha k_0 |c|\mbox{sgn}F\over L'(k_0)}\quad\quad \Longrightarrow
\nonumber\\
\mbox{sgn}(\mbox{Im} k) = \mbox{sgn}(cF)\,\mbox{sgn}[ck_0 L'(k_0)], \label{cab8}
\end{eqnarray} 
as $\alpha\to 0+$. We know that the contour $C$ in Eq.\ (\ref{cab4}) lies in the
upper (resp.\ lower) half plane provided $cFx>0$ (resp.\ $cFx<0$). Therefore, the
poles whose residues count must satisfy $cFx$Im$k>0$. Then Eq.\ (\ref{cab8}) implies
that we should count poles satisfying $xck_0 L'(k_0)>0$. The physical meaning of this
criterion becomes clear if we calculate the group velocity corresponding to mode
$k_0$. $L(k)=0$ yields $(kc)^2= [\omega(k)]^2 = A + 4 \sin^2 (k/2)$. Then $2
\omega(k) \omega'(k)= 4\sin(k/2)\cos(k/2) = L'(k) + 2kc^2$. Thus $v_g \equiv
\omega'(k)$ obeys
\begin{eqnarray} 
v_g = c + {L'(k)\over 2kc} \Longrightarrow 
\mbox{sgn} (v_g - c) = \mbox{sgn}[ck L'(k)], \label{cab9}
\end{eqnarray}
and we observe that the poles whose residues contribute to the solution satisfy $(v_g
- c)\, x>0$. This was the criterion used by Atkinson and Cabrera \cite{atk65}: All
modes with $v_g>c$ must appear ahead of the wave front ($x>0$), all those with $v_g<c$
must appear behind ($x<0$). See Figure \ref{figura3}.

\begin{figure}
\begin{center}
\includegraphics[width=8cm]{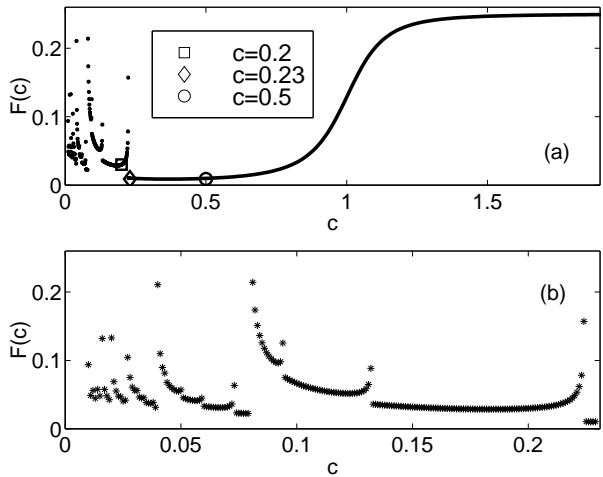} 
\end{center}
\caption{(a) $F(c)$ for $\alpha=0$, $A=0.25$ as computed from formula (\ref{cab9}),
(b) Zoom in the region of resonances for $c$ small.}
\label{figura5}
\end{figure}

For $c>0$ and $\alpha=0$, the condition (\ref{cab7}) becomes \cite{atk65} 
\begin{eqnarray} 
{F\over A} = \sum_{L(k)=0, k>0} {2A\over k\, |L'(k)|}. \label{cab10}
\end{eqnarray} 
This formula follows straightforwardly from the fact that $L(k)$ and $k\, L'(k)$ are
even functions of (real) $k$ and symmetry considerations. Notice that our assumption
$c>0$ has yielded $F>0$. The relation $F(c)$ given by (\ref{cab10}) is plotted in
Figure \ref{figura5} for a value $A=0.25$ (see also Fig.\ 3 of
Ref.~\onlinecite{atk65}).  For a given value of the external force $F$, there may be
several values of admissible velocities $c$, each corresponding to a different wave
front profile. Thus different families of wave fronts (not all of them stable) may
coexist for the same value of $F$. The function $F(c)$ presents different vertical
asymptotes at positive values $c_i$, $c_1>c_2>\ldots$, where both $L(k)$ and $L'(k)$
and vanish for positive $k$. We have $c_n\sim \sqrt{A}/(2\pi n)$ as $n\to \infty$, so
that the vertical asymptotes accumulate at $c=0$ as suggested by Fig. \ref{figura3a}.
In fact, the velocity can be eliminated from the two conditions $L(k)= L'(k)= 0$
yielding $A+ 2\, (1-\cos k) = k\sin k$. For large values of $k$, this gives $k \sim
2\pi n + A/(2\pi n)$, as the integer $n\to\infty$. The condition $L'(k)=0$ then
yields the previous formula for $c_n$.

The range of physical interest corresponds to $c_1<c<1$ (wave front velocities larger
than the largest resonant velocity but smaller than the sound speed \cite{atk65}). Let
$(c_m, F(c_m))$ be the minimum of the first and fastest branch of $F(c)$. Then $F_{cd}
= F(c_m)$ yields the dynamic Peierls force, under which the `physical' branch of {\em
stable} wave front profiles ceases to exist. This force is smaller than the static
Peierls force, $F_c= A^{{3\over 2}}\, (A+4)^{-{1\over 2}}$, for the piecewise linear
model \cite{atk65}.

\begin{figure}
\begin{center}
\includegraphics[width=8cm]{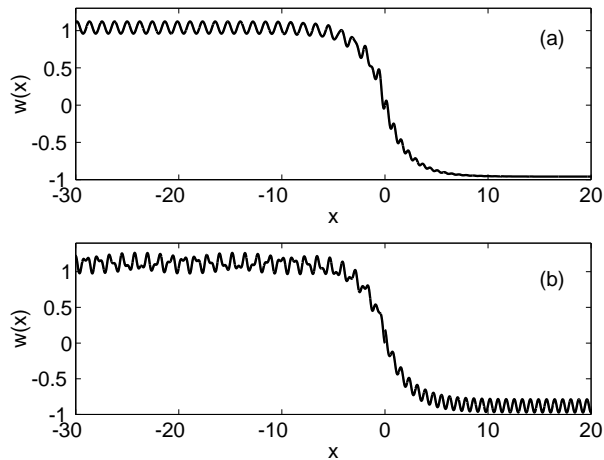} 
\end{center}
\caption{Wave front profiles for $A=0.25$, $\alpha=0$ and: (a) $c=0.23$, (b) $c=0.2$.}
\label{figura6}
\end{figure}

Using the radiation condition $(v_g -c)\, x >0$, we have plotted in Figures
\ref{figura1} and \ref{figura6} several wave front profiles. Figure \ref{figura1}(a)
shows a  wave front profile for $\alpha=0$, $A=0.25$ and $F=0.009$. The wave velocity
is $c=0.5$. The profile has been numerically approximated by computing the contour
integral (\ref{cab5}) as the series of residues truncated to a few terms. It is
interesting to observe that the right tail of the wave front decays fast to $U_1(F/A)$
whereas the left tail oscillates about $U_3(F/A)$ with uniform amplitude. The reason
for this behavior is that for the above parameter values, there are two real zeros of
$L(k)$ at $\pm k_0$ with $k_0 L'(k_0)>0$ that contribute a non-decaying oscillation
to the left tail. An infinitesimal amount of friction would dampen these oscillations
by contributing a multiplicative factor exp$\{- [\alpha k_0 c/L'(k_0)]\}$ to their
amplitude. Figure \ref{figura6}(a) shows a wave front profile for $c=0.23$. Notice
the decaying small oscillation in the right tail. We have still two real poles, but
now $c=0.23$ is placed at the left of the minimum in the first branch of $F(c)$; see
Figure \ref{figura5}(a). Figure \ref{figura6}(b) shows a wave front profile for
$c=0.2$, in the second branch of $F(c)$, past the first resonance. Now we have three
pairs of real poles. Two of them contribute to the oscillation in the left tail,
the other one produces the oscillation in the right tail.

\begin{figure}
\begin{center}
\includegraphics[width=8cm]{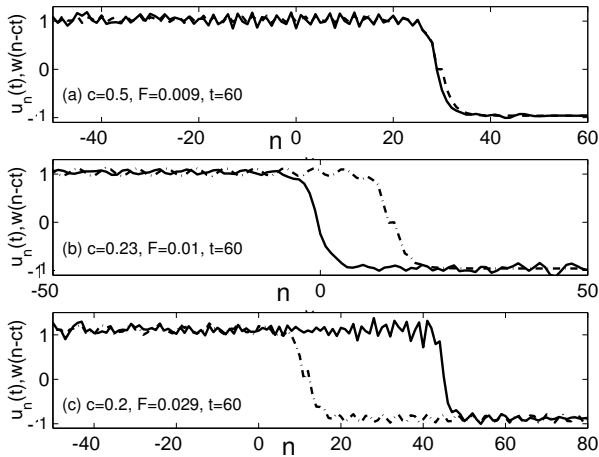} 
\end{center}
\caption{Dynamical stability when $A=0.25$, $\alpha=0$. We compare $u_n(t)$ (solid
line) to $w(n-ct)$ (dot-dashed line) for (a) $c=0.5$, (b) $c=0.23$, (c) $c=0.2$. }
\label{figura7}
\end{figure}

The dynamical stability of the constructed wave front solutions can be
numerically checked by using their computed profile as initial data to solve Eq.\
(\ref{cab1}) with $\alpha=0$ \cite{numerical}. The results are compared at a fixed
time $t= 60$ to the expected configuration $w(n-ct)$ in Figure \ref{figura7}. The
choice $c=0.5$ seems to produce a stable wave front. The choice $c=0.23$, (still on
the first and fastest branch of $F(c)$ but to the left of its minimum), evolves
towards a static front. The choice $c=0.2$ (on the second branch of $F(c)$), evolves
towards a wave front moving faster than expected, with a speed on the first branch of
$F(c)$. Thus our numerical results seem to indicate that stable wave fronts have
velocities on the first and fastest branch of $F(c)$ with $F'(c)>0$, to the left of
the minimum speed on this branch, $c_m>0$. Then $F_{cd}= F(c_m)$, and stable wave
fronts with $v'(1)<0$ have speeds larger or equal than $c_m$.

We have found wave front profiles with oscillatory tails that seem stable under small
disturbances. One question that comes to mind is whether these profiles occur in
models with smooth nonlinearities. The answer is yes: See an explicit construction
in Appendix A. 

\subsection{Overdamped limit: $\alpha=\infty$}

\begin{figure}
\begin{center}
\includegraphics[width=8cm]{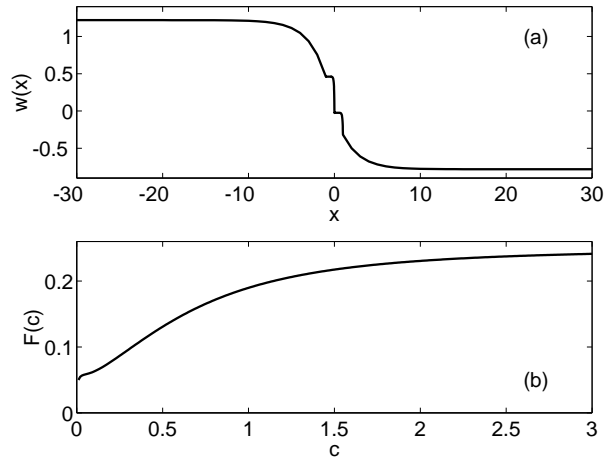}
\end{center}
\caption{Overdamped limit when $A=0.25$. (a) Wave front profile for
$c=0.02$. (b) $F(c)$ as computed from formula (\ref{cab7}).}
\label{figura8}
\end{figure}

The results in the overdamped limit $m=0$ are consistent with previous work
\cite{cb01,kin01,fat98}: there are one wave front profile and one $c$ for each fixed
$F$ above a threshold, $F_c$. Wave front profiles are monotone, and they resemble
staircases for $c$ small. See Figure \ref{figura8}.

\subsection{Finite damping: $\alpha>0$}
\begin{figure}
\begin{center}
\includegraphics[width=8cm]{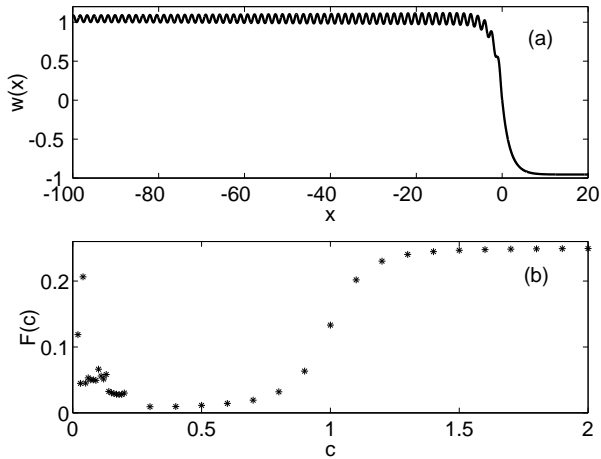}
\end{center}
\caption{Results for $A=0.25$, $\alpha=0.01$ (a) Wave front profile for
$c=0.5$, (b) $F(c)$ as computed from formula (\ref{cab7}).}
\label{figura10}
\end{figure}

\begin{figure}
\begin{center}
\includegraphics[width=8cm]{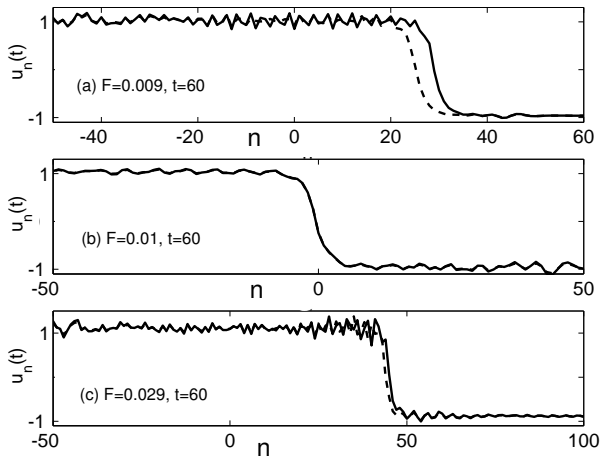}
\end{center}
\caption{Comparison between $u_n(t)$ calculated for $\alpha=0$ for the same values as
in Fig. \protect\ref{figura7} (solid line) and $u_n(t)$ for $\alpha=0.01$ (dot-dashed
line). (a) $F=0.009$, (b) $F=0.01$, (c) $F=0.029$. }
\label{figura7b}
\end{figure}

The results for finite damping interpolate between the conservative and overdamped
cases. For small $\alpha$, the function $F(c)$ and the wave front profiles are non
monotone although their oscillations decay as $n\to\pm\infty$; see Fig.
\ref{figura10}. Fig. \ref{figura7b} shows a comparison between $u_n(t)$ for $\alpha=
0$ (for the same values as in Fig. \ref{figura7}). We observe that, for this small
damping, the corresponding wave fronts have the same stability properties as in the
conservative case: dynamically stable for $c>c_m$, and unstable for $c<c_m$. Moreover
there are dynamic and static Peierls stresses which are different from each other, as
in the case $\alpha =0$.

\begin{figure}
\begin{center}
\includegraphics[width=8cm]{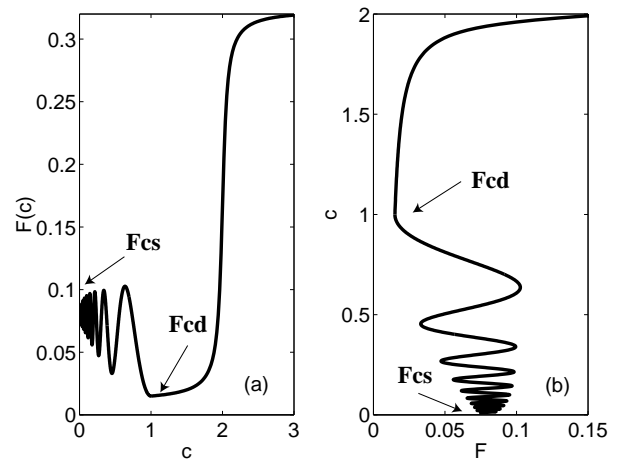}
\end{center}
\caption{(a) Schematic function $F(c)$ for small $\alpha>0$ showing infinitely many
maxima accumulating at $c=0$ and $F=F_{cs}$. (b) The bifurcation diagram of wave
front velocity versus $F$: there are infinitely many limit points (saddle-node
bifurcations) corresponding to the extrema of $F(c)$ in the interval $F_{cd} < F <
F_{cs}$. }
\label{figura14}
\end{figure}

It is interesting to ascertain the shape of the function $F(c)$ because its inverse
function constitutes a bifurcation diagram clarifying wave front depinning. In the
conservative case, $F(c)$ has infinitely many vertical asymptotes (resonances)
accumulating at $c=0$. As $\alpha\to 0+$, the conditions $L(k,\alpha)=L_k(k,\alpha)=
0$ yield
\begin{eqnarray} 
k &\sim& k_0 +\left| {\alpha k_0 c\over \cos k_0 - c^2} \right|^{{1\over 2}}
e^{i{\pi\over 4}\chi + i\pi l}, \quad l=0,1, \label{cab11}\\
\chi &=& \mbox{sgn}[(\cos k_0 - c^2) k_0 F]. \label{cab12}
\end{eqnarray} 
This formula shows how the double poles $k=k_0$ of the conservative case split when
an infinitesimal friction is present. Then the vertical asymptotes of $F(c)$ at
$c=c_n$ give rise to local maxima of $F(c)$ for small $\alpha >0$. These maxima are
hard to resolve numerically (see Fig. \ref{figura10}(b)), but they can be
approximately calculated as follows. Only poles with positive imaginary part
contribute to the sum in Eq.\ (\ref{cab7}). For these poles, the exponential factor
in Eq.\ (\ref{cab11}) is $(\chi + i)/\sqrt{2}$, and their contribution to the sum in
Eq.\ (\ref{cab7}) is approximately given by 
\begin{eqnarray} 
{(2\alpha)^{-{1\over 2}}\, A\, (\chi -i)\over k_0\,\mbox{sgn} (\cos k_0 - c^2)\, |k_0
c\, (\cos k_0 - c^2)|^{{1\over 2}}Ê} . \nonumber
\end{eqnarray} 
To this expression, we should add its complex conjugate, a contribution to the sum in
Eq.\ (\ref{cab7}) due to the pole $-k_0$. If we keep only these contributions in Eq.\
(\ref{cab7}), thereby assuming that the considered maximum of $F(c)$ is large, $F_M
\gg 1$, we obtain
\begin{eqnarray} 
|F_M| \sim {\sqrt{2} A^2\alpha^{-{1\over 2}}\over |k_0|^{{3\over 2}}\, |c\, (\cos
k_0 - c^2)|^{{1\over 2}}Ê} . \label{cab13}
\end{eqnarray} 
Now, in the conservative case, $c_n\sim \sqrt{A}/(2\pi n)$ and $k_0 \sim 2\pi n +
A/(2Ê\pi n)$ as the integer $n\to\infty$. Then the right side of Eq.\ (\ref{cab13})
becomes proportional to $c_n$. When $n$ is so large that $c_n$ is no longer large
compared to $\sqrt{\alpha}$, other terms of the sum contribute appreciably to $F$ in
the formula (\ref{cab7}). We conjecture that these contributions add to $F_{cs}$,
\begin{eqnarray} 
|F_M| - F_{cs} \sim \sqrt{{2\over\alpha}}\, A^{{5\over 4}}\, c_n, & c_n \sim
{\sqrt{A} \over 2\pi nÊ} , \label{cab14}
\end{eqnarray} 
so that the maxima of $F(c)$ accumulate near $c=0$ as the integer $n\to\infty$. We
have depicted schematically the resulting $F(c)$ and the bifurcation diagram of $c$
versus $F$ in Fig. \ref{figura14}. Corresponding to the infinitely many local extrema
in Fig.Ê\ref{figura14}(a), there are infinitely many limit points (saddle-node
bifurcations) in Fig. \ref{figura14}(b). Our numerical results indicate that only the
branch of wave fronts with larger velocities in the physical interval $c_m < c$ and
$|F| >F_{cd}$ are stable. This can be understood from the factorization theorems in
Ref.Ê\onlinecite{ioo80}. According to these theorems, one eigenvalue of the linear
stability problem corresponding to the solution branches in the bifurcation diagram
changes sign at limit points. If we use that the branch of wave fronts with larger
velocity is stable, the branch that coalesces with it at the limit point $F=F_{cd}$
is unstable. This branch coalesces with another one at another limit point with
larger $F$, and, there, a different eigenvalue of the linear stability problem changes
sign from negative to positive. If this is so, the new branch is also unstable and all
other wave front solution branches in Fig. \ref{figura14}(b) could also be unstable.

\begin{figure}
\begin{center}
\includegraphics[width=8cm]{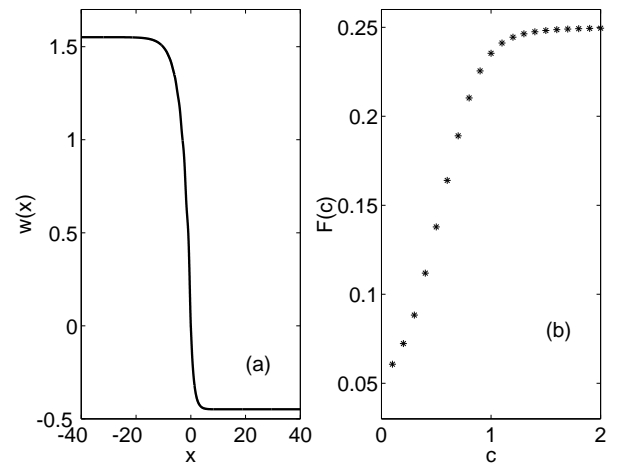}
\end{center}
\caption{Results for $A=0.25$, $\alpha= 1$. (a) Wave front profile for
$c=0.5$, (b) $F(c)$ as computed from formula (\ref{cab7}). }
\label{figura9}
\end{figure}

For larger values of $\alpha$, the wave front profiles become monotone, the
oscillation amplitudes in $F(c)$ decrease and become difficult to appreciate; see Fig.
\ref{figura9}. The transition from one parameter range to the other one occurs when
the contribution from poles with small imaginary part in Eq.\ (\ref{cab7}) becomes
relevant.

\section{Wavy wave fronts for generic cubic nonlinearities}
\label{sec:generic}
For generic smooth cubic nonlinearities $g(u)$, we cannot construct the wave front
profiles by using contour integrals. However, we can extend our previous theory of
the active points \cite{cb01} for threshold phenomena to the case of finite damping.
Thus we shall present numerical results for (relatively) large $A$ showing that wave
fronts are similar to those for piecewise linear $g(u)$. Near $F_c$, we shall use the
theory of active points to interpret numerical results. 

\subsection{Numerical results}

\begin{figure}
\begin{center}
\includegraphics[width=8cm]{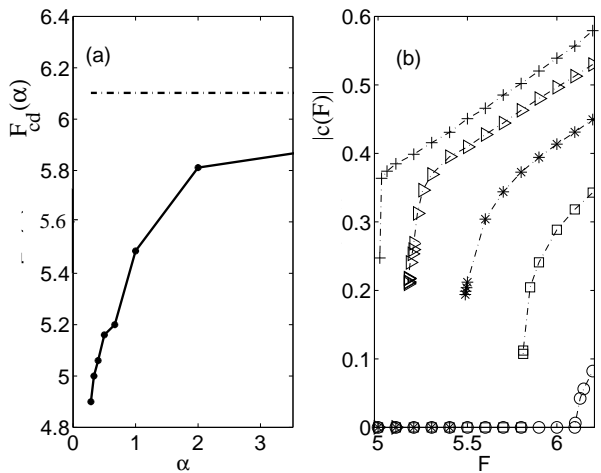}
\end{center}
\caption{Results for the Frenkel-Kontorova model with $A=10$: (a) Dynamic (solid line)
versus static (dot-dashed line) thresholds as functions of $\alpha$; (b) numerical
velocities as functions of $F$ for decreasing values of $\alpha$: 3 (circles), 1.5
(squares), 1 (asterisks), 0.7 (triangles), 0.57 (crosses). }
\label{figura11}
\end{figure}

\begin{figure}
\begin{center}
\includegraphics[width=8cm]{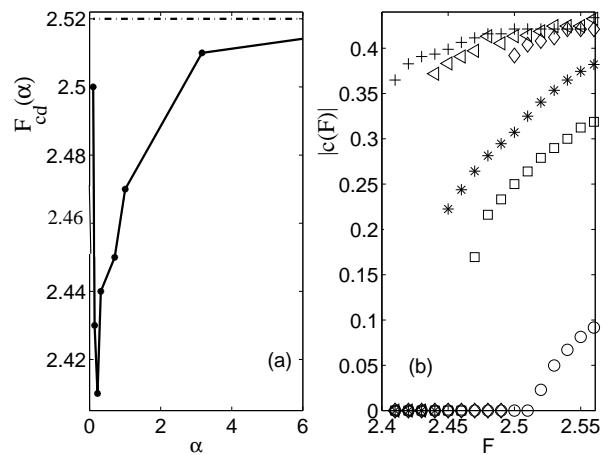}
\end{center}
\caption{Same as Fig. \protect\ref{figura11} for the model with a quartic potential.
In (b), the values of $\alpha$ are: 3 (circles), 1 (squares), 0.7 (asterisks), 0.3
(triangles), 0.2 (crosses), 0.1 (diamonds). }
\label{figura12}
\end{figure}

\begin{figure}
\begin{center}
\includegraphics[width=8cm]{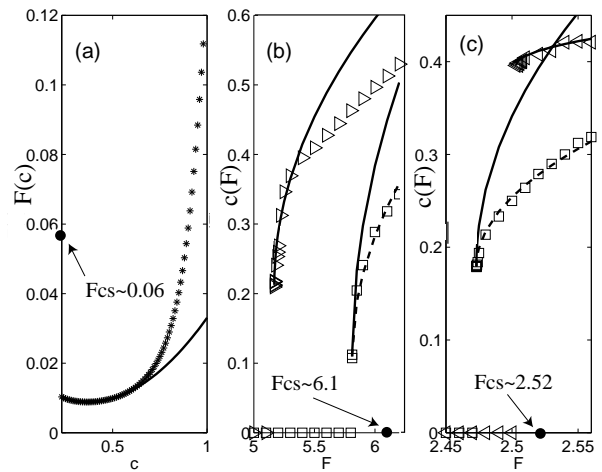}
\end{center}
\caption{(a) Function $F(c)$ for the piecewise linear $g(u)$ with $A=0.25$ and
$\alpha=0$. (b) Velocity versus applied stress for $g(u)=\sin u$ with $A=10$ and
$\alpha=1.4$ (squares) and $\alpha=0.57$ (triangles). (c) Velocity versus applied
stress for $g(u)= u(u^2-1)$ with $A=10$ and $\alpha=1.0$ (squares) and $\alpha=0.1$
(triangles).}
\label{figura19}
\end{figure}

Figures \ref{figura11} and \ref{figura12} show the dynamic and static Peierls
stresses and the wave front velocity for the damped FK and cubic models,
respectively. We have constructed the traveling wave fronts by solving numerically
Eq.\ (\ref{cab1}) in large lattices. For $F> F_{cs}$, we choose as initial data a
static step-like profile: $u_n(0)=U^{(1)}(F/A)$ if $n \geq 0$, $u_n(0)=U^{(3)}(F/A)$
if $n<0$, $u_n'(0)=0$ $\forall n$. We use the stable zeros $U^{(1)}(F/A)$ and
$U^{(3)}(F/A)$ as boundary conditions for large $|n|$. The numerical solution
$u_n(t)$ evolves very fast to a traveling wave $u_n(t)=w(n-ct)$ with a fixed constant
value for the speed $c$. For $F$ below the static threshold, we choose as initial data
the traveling solutions already found. As boundary condition, we use again
$U^{(1)}(F/A)$ and $U^{(3)}(F/A)$. The numerical solutions $u_n(t)$ evolve to a
traveling wave $u_n(t)=w(n-ct)$ with a profile and speed adjusted to the new value of
$F$, provided $F$ is larger than the dynamical threshold $F_{cd}$. Below that value,
the waves are pinned. The behavior of $c$ near $F_{cd}$ can be guessed from the known
fact that the function $F(c)$ has a minimum $F_{cd}$ on its fastest branch at $c=c_m$
for piecewise linear $g(u)$; see Fig.~\ref{figura5}. Near this minimum, $F \sim
F_{cd} + \gamma\, (c-c_m)^2$ with $\gamma>0$, as indicated in Fig.~\ref{figura19}(a).
This yields $(c-c_m) \sim [(F-F_{cd})/\gamma]^{{1\over 2}}$, a scaling that can be
seen in Figs. \ref{figura19}(b) and (c), corresponding to smooth $g(u)$. The number
$\gamma$ can be fitted by taking careful numerical measurements near $F_{cd}$. This
seems to provide a good fitting over an interval of stresses that increases as
$\alpha$ decreases. For larger friction values, taking values of $F$ farther from
$F_{cd}$ produces a better fit to a scaling with the same exponent 1/2 but with
different $\gamma$; see the squares in Figs. \ref{figura19}(b) and (c). 
 
\begin{figure}
\begin{center}
\includegraphics[width=8cm]{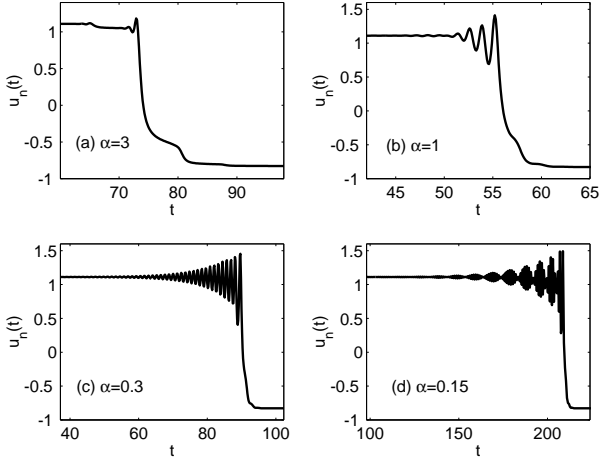}
\end{center}
\caption{Wave front profiles for the quartic potential as seen from the trajectory of
a single point $u_n(t)$: (a) $\alpha=3$, (b) $\alpha=1.5$, (c) $\alpha=0.3$, (d)
$\alpha=0.15$. }
\label{figura13}
\end{figure}

Our numerical measurements of the speeds near $F_{cd}$ seem to indicate that (except
in the overdamped limit $m=0$) there is a critical non zero speed $c_m>0$ below which
front propagation cannot be sustained. In the coexistence region, $F_{cd}<F<F_{cs}$,
shown in Figs. \ref{figura11} and \ref{figura12}, both the traveling wave fronts and
the static wave fronts are dynamically stable. The wave front profiles for different
damping values and the cubic $g(u)$ are depicted in Fig. \ref{figura13}. These
profiles oscillate more and more as the damping coefficient decreases. For
sufficiently large $\alpha$, the wave front profiles are monotone and become similar
to those calculated in the overdamped limit \cite{cb01}. $F_{cd}$ and $F_{cs}$ are
almost equal. Similarly, as $A$ decreases and we approach the continuum limit, the
gap between static and dynamical thresholds is difficult to appreciate.

There is an important difference between models with a periodic nonlinearity such as
FK and models with a cubic $g(u)$. In both cases, wave fronts can be constructed
numerically for sufficiently large values of the damping. For $A=10$, wave fronts of
the cubic model can be numerically found at least for $\alpha\geq 0.08$, whereas those
of the FK model are found easily for larger damping, $\alpha >0.5$. For smaller values
of the damping, the amplitude of the wave front oscillatory tails becomes so large
that the FK wave front profile tends to jump between different periods of the
nonlinearity: moving staircases are thus generated.

\subsection{Active point theory}
\label{sec:onset}
To get approximate formulas for the wave front profile and velocity in the strongly
discrete limit, $A\gg 1$, we can resort to the active point theory \cite{cb01}. In
this limit, there is one active point, say $u_0(t)$, and all others obey either $u_n
\sim U_1(F/A)$ (for $n>0$) or $u_n \sim U_3(F/A)$ (for $n<0$). We assume that the wave
front we will construct has $F>0$, $c\geq 0$ and $w'(0)<0$, as in the previous
Section. According to Eq.\ (\ref{cab1}), the active point satisfies the approximate
equation:
\begin{eqnarray}
{d^2u_{0}\over dt^2} + \alpha\, {du_{0}\over dt}\approx U_1\left({F\over A}
\right) + U_3\left({F\over A} \right) \nonumber\\
-2 u_0 - A\, g(u_0) + F . \label{a1}
\end{eqnarray}
This equation has three stationary solutions for $F<F_{cs}$, $z_1(F_c/A)< z_2(F_c/A)
< z_3(F_c/A)$, ($z_1$ and $z_3$ are stable and $z_2$ is unstable), and only one stable
stationary solution, $z_3$, for $F>F_{cs}$. The critical field $F_c= F_{cs}$ is such
that the expansion of the right hand side of (\ref{a1}) about the two coalescing
stationary solutions has zero linear term, $2 +A g'(u_0)=0$, and 
\begin{equation}
2 u_0 + A \, g(u_0) \sim  U_1\left({F_c\over A}
\right) + U_3\left({F_c\over A} \right) + F_c .
\label{a2}
\end{equation}
For $F$ sligthly above $F_c$, $u_0(t)= u_0(A,F_c) + v_0(t)$ obeys the following
equation:
\begin{eqnarray}
{d^2v_{0}\over dt^2} &+& \alpha\, {dv_{0}\over dt}\approx a\, (F-F_c) + b\, v_{0}^{2},
\label{a3}\\
a &=& 1 + {1\over A\, g'(U_{1}(F_{c}/A))} + {1\over A\,
g'(U_{3}(F_{c}/A))} >0,\quad\quad \label{a4}\\
b &=& -{A\over 2}\, g''(u_0) >0, \label{a5}
\end{eqnarray}
where we have used $2+Ag'(u_0)=0$, (\ref{a2}) and ignored higher order terms. This
equation has two distinguished limits, $\alpha\ll (F-F_c)^{{1\over 4}}\ll 1$ and
$\alpha\gg (F-F_c)^{{1\over 4}}$. In the latter case, we can ignore the inertia in
Eq.\ (\ref{a3}). The resulting reduced equation is exactly that analyzed in Ref.
\onlinecite{cb01} except for a trivial rescaling of the time. The solution of that
equation blows up at times $(t-t_0)\sim \pm\pi\alpha/[2\sqrt{ab\, (F-F_{c})}]$ ($t_0$
is an arbitrary constant). Then the wave front velocity is approximately given by the
reciprocal of the interval between two consecutive blow up times, namely,
\begin{eqnarray}
c = {\sqrt{ab\, (F-F_c)}\over \pi\alpha}\,. \label{a6}
\end{eqnarray}
After blow up, the wave front profile is reconstructed by inserting an inner layer,
in which $u_0(t)$ obeys Eq.\ (\ref{a1}) with $F=F_c$, and it jumps from a neighborhood
of $z_1(F_c/A)$ to $z_3(F_c/A)$ \cite{cb01}.

If $\alpha\ll (F-F_c)^{{1\over 4}}\ll 1$, we can ignore friction in Eq.\ (\ref{a3})
thereby obtaining a conservative dynamical system, Eq.\ (\ref{a3}) with $\alpha=0$,
as our reduced equation. Its trajectories also blow up and the wave front velocity
can be straightforwardly calculated as
\begin{eqnarray}
c = {\left({ab\, (F-F_c)\over 3}\right)^{{1\over 4}}\over 2\sqrt{2}\,
K\left({1\over \sqrt{2}} \right)} = {\sqrt{2\pi}\, \left({ab\, (F-F_c)\over 3}\right)^{{1\over 4}}\over 
\left[\Gamma\left({1\over 4} \right)\right]^2}\,, \label{a7}
\end{eqnarray}
where $K(1/\sqrt{2})= 1.854075$ is the complete elliptic integral of the first kind
with module $k=1/\sqrt{2}$ \cite{gra80}. Contrary to the overdamped case, a
consistent inner layer connecting blowing up trajectories of the reduced equation and
trajectories of Eq.\ (\ref{a1}) for $F=F_c$ and $\alpha=0$ does not exist. This
points out to a breakdown of the active point theory as $\alpha\to 0+$, which is
consistent with our conjectured bifurcation diagram in Fig. \ref{figura14}(b) for the
piecewise linear model. Assuming that Fig. \ref{figura14}(b) is also the bifurcation
diagram of models with smooth nonlinearities, a succession of infinitely many
saddle-node bifurcations (that accumulate at $c=0$, $F=F_{cs}$) connect the branch of
stable wave fronts and that of stationary front solutions between $F_{cd}$ and
$F_{cs}$. If this is the case, seeking a description in terms of standard normal
forms as (\ref{a3}) and scaling is rather hopeless. Not surprisingly, the suspicious
scaling (\ref{a7}) is hard to check numerically (the stable branch of moving wave
fronts ends at $F= F_{cd}<F_c$), whereas the scaling (\ref{a6}) can be easily checked
($F_{cd}\approx F_c$ in the overdamped limit) \cite{cb01}.

\begin{figure}
\begin{center}
\includegraphics[width=8cm]{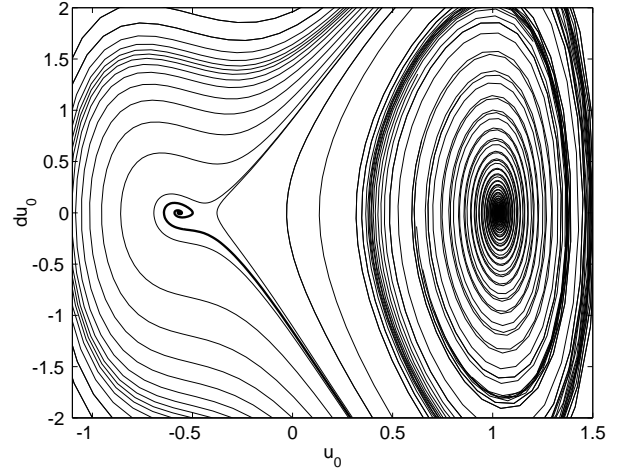}
\end{center}
\caption{Phase plane of Eq.\ (\ref{a1}) with the cubic nonlinearity for $A=10$,
$\alpha=0.5$ and $F=2.45$.}
\label{figura15}
\end{figure}

Let us now come back to Eq.\ (\ref{a1}) to explain the coexistence of moving and
stationary wave fronts for $F_{cd}< F < F_c$. In Fig. \ref{figura15}, we have depicted
the phase plane associated to Eq.\ (\ref{a1}) for the cubic nonlinearity $g(u)= u\,
(u^2 -1)$. The profiles in Fig. \ref{figura13} correspond to one active point jumping
from a neighborhood of $U_1(F/A)$ to a neighborhood of $U_3(F/A)$. 
\begin{figure}
\begin{center}
\includegraphics[width=8cm]{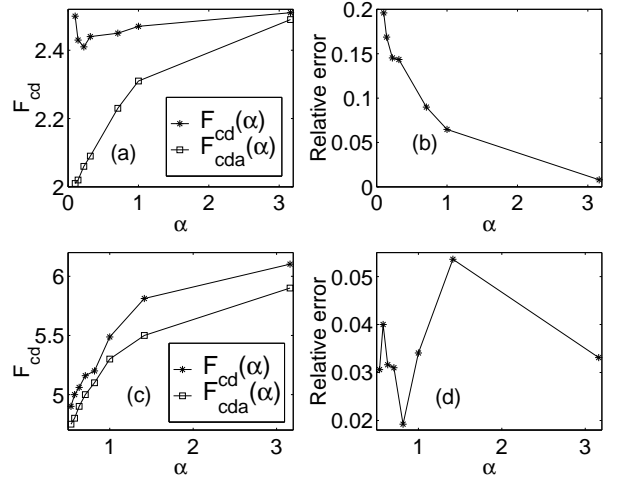}
\end{center}
\caption{Comparison of $F_{cda}\equiv F_{c2}$ and $F_{cd}$ as functions of the damping
$\alpha$ for $A=10$ and: (a) the cubic nonlinearity, (b) its relative error, (c) the
FK nonlinearity, and (d) its relative error.}
\label{figura18}
\end{figure} 
The right side of Eq.\ (\ref{a1}) has three zeros in the interval $[U_1(F/A),
U_3(F/A)]$ for $F<F_{c_1}$, corresponding to two stable spiral points or centers
$(z_i(F/A),0)$, $i=1,3$, and one saddle point $(z_2(F/A),0)$. At $F=F_{c_1}$ two of
the zeros coalesce and for $F>F_{c_1}$ only the spiral point $(z_3(F/A),0)$ remains.
For $\alpha >0$, there is a new critical value of $F$, $F_{c_2}$, such that the
initial datum $(U_1(F/A),0)$, which is close to the spiral point $(z_1(F/A),0)$, may
evolve to the other stable spiral point, $(z_3(F/A),0)$, for $F\in(F_{c_2},F_{c_1})$;
see Fig. \ref{figura15}. The trajectory leaving a neighborhood of $(z_1(F/A),0)$ and
entering $(z_3(F/A),0)$ defines the wave front profile. For $F<F_{c2}$, the initial
datum $(U_1(F/A),0)$ evolves toward $(z_1(F/A),0)$ and no wave front is generated.
Thus $F_{c2}$ yields an approximation to the dynamical Peierls stress, $F_{cd}$. Fig.
\ref{figura18} compares $F_{c2}$ to $F_{cd}$, which has been calculated numerically
by solving the complete system (\ref{cab1}). Notice that our approximation worsens as
$\alpha$ decreases towards zero indicating break down of the one active point
approximation. We shall explain below why this is so.

\begin{figure}
\begin{center}
\includegraphics[width=8cm]{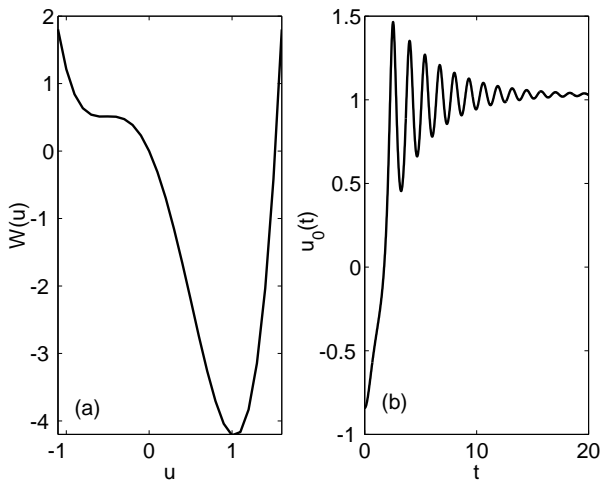}
\end{center}
\caption{(a) Potential energy $W(u)$ corresponding to the cubic nonlinearity
$h(u)=2u_0 +Ag(u_0) -F - U_1(F/A) - U_3(F/A)$, with $A=10$ and $F=2.45$. (b)
Trajectories of Eq.\ (\ref{a1}) with initial condition $(U_1(F/A),0)$, joining
$(z_1(F/A),0)$ and $(z_3(F/A),0)$ for $A=10$, $\alpha=0.5$ and $F=2.45$. }
\label{figura16}
\end{figure}

\begin{figure}
\begin{center}
\includegraphics[width=8cm]{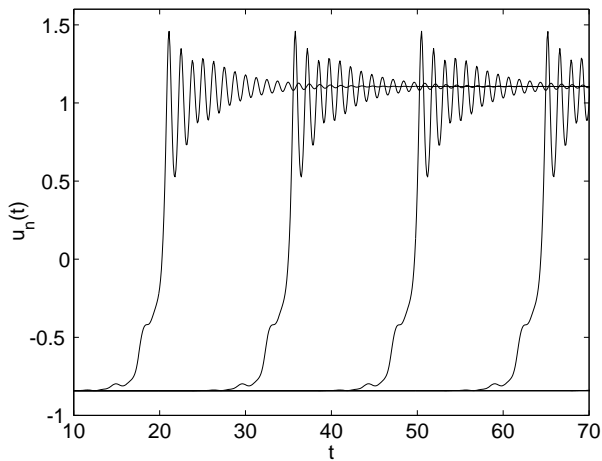}
\end{center}
\caption{Numerical solutions $u_n(t)$ of Eq.\ (\ref{cab1}) with the cubic nonlinearity
for $A=10$, $\alpha=0.5$ and $F=5.9$.}
\label{figura17}
\end{figure}

The dynamic critical Peierls stress can be intuitively explained as follows. The
potential energy associated to the nonlinearity $h(u)=2u_0 +Ag(u_0) -F - U_1(F/A) -
U_3(F/A)$ on the right side of Eq.\ (\ref{a1}) is depicted in Fig. \ref{figura16}(a).
An initial condition $(U_1(F/A),0)$ is close to the left minimum of $W(u_0)$. If the
energy corresponding to the initial condition is slightly higher than that of the left
minimum, the solution of (\ref{a1}) evolves toward it because of friction. On the
other hand, larger energies cannot be dissipated by the friction term and the
trajectory surpasses the maximum of $W(u_0)$ with nonzero velocity. Then the
trajectory falls in the basin of the right minimum of the potential, performing a
damped oscillation about it before reaching $z_3(F/A),0)$. Once $u_0(t)$ has reached
a neighborhood of the second spiral point, $u_{-1}(t)$ takes its place and performes a
similar motion. The resulting trajectory is depicted in Fig. \ref{figura16}(b). The
solution of the complete system (\ref{cab1}) is shown in Fig. \ref{figura17}. The wave
front velocity is approximately given by the reciprocal of the time $u_0(t)$ takes to
jump from $z_1(F/A)$ to $z_3(F/A)$. Notice that the oscillations about $z_3(F/A)$
persist a longer time as $\alpha$ decreases and may have finite amplitude. Then we
cannot approximate sufficiently well $u_1(t)$ by the constant value $U_3(F/A)$ for
small values of the friction and the one active point approximation breaks down. This
explains the discrepancies in Fig. \ref{figura18} for small $\alpha$. Since the
difference $[U_3(F/A)-U_1(F/A)]$ is larger for the FK than for the cubic
nonlinearity, the approximation $u_1(t)\approx U_3(F/A)$ is better for the FK
nonlinearity as shown in Fig. \ref{figura18}.

\section{Discussion}
\label{sec:conclusions}
We have studied wave front solutions in chains of nonlinear oscillators with inertia
and damping. Two analytical methods have been used to construct the wave fronts and
their velocities. For piecewise linear models, exact formulas can be found for the
wave fronts and the relation between their velocity and the applied stress $F$ as
Atkinson and Cabrera did already in 1965 for the conservative case \cite{atk65}.
Different from these authors, we have also studied the cases of finite and infinite
damping. We have depicted the resulting wave front profiles for all damping values and
found that they may have oscillatory tails. For zero damping, these tails oscillate
with non-decaying amplitude as $n\to\infty$, which means that a new definition of wave
front is needed \cite{definition}. We have shown numerically that non-monotone wave
fronts with oscillatory tails ({\em wavy wave fronts}) may be stable for certain
intervals of applied stresses. We have found stable moving wave fronts for
$|F|>F_{cd}>0$ (the dynamic Peierls stress), and these fronts coexist with stable
static wave fronts for an interval $F_{cd} < |F|<F_{cs}$ ($F_{cs}$ is the static
Peierls stress; static wave fronts exist for $0 \leq |F|<F_{cs}$). 

We have also conjectured that the global bifurcation diagram for wave front depinning
in the presence of inertia and damping is generically as in Fig. \ref{figura14}(b).
Then there are infinitely many saddle-node bifurcations between wave front branches in
the interval $F_{cd}<|F| < F_{cs}$, accumulating at $c=0$ and $F=F_{cs}$, with
$F_{cd} >0$. The basis for this conjecture are our results for the piecewise linear
model, that we think are generic also for models with smooth nonlinearities given our
numerical results for them. The function $F(c)$ in the piecewise linear model has
infinitely many vertical asymptotes (resonances) for zero damping that accumulate as
$c\to 0$ and (we suppose) $F\to F_{cs}$. These resonances become local maxima of
$F(c)$ as a small damping is added to the model. Maxima and minima of $F(c)$ are
limit points (saddle-node bifurcations) between branches of wave front solutions in
the diagram of Fig. \ref{figura14}(b).

In the literature \cite{sch79,fla99}, wave fronts with monotone profiles have been
constructed for undamped models. For example, Schmidt \cite{sch79} showed that $w(x)
= \tanh x$, $x= n+t/2$, is a wave front solution for Eq.\ (\ref{cab1}) with
$\alpha=F=0$, $A= 1$ and a potential $V(u)= - 3u^2/4 - u^4/8 - (\sinh 1)^{-2}\,
\ln(1- u^2 \tanh^2 1)$, with $g(u)=V'(u)$. For this model, $F_{cs}>0$ \cite{fla99}.
Solving numerically this model for $F\neq 0$, wave fronts with one oscillatory tail
on $x>0$ are found for $F>0$, similar to those in Fig. \ref{figura1}(b). For $F<0$,
wave fronts with one oscillatory tail on $x<0$ are found instead; see Fig.
\ref{figura1}(c). Both types of fronts have $|c|>1/2$. This shows that wavy wave
fronts are generic and that Schmidt's monotone wave front is a non-generic limiting
case separating branches of wavy wave fronts and corresponding to $F_{cd}=0$. A
possible bifurcation diagram $|c|$ vs.\ $F$ for this example could be that in Fig.
\ref{figura14}(b) with $F_{cs}>0$, $F_{cd}=0$ and $|c_m|=1/2$: numerical solution of
the model seems to be consistent with a bifurcation diagram having a limit point at
$F=0$, $|c|=1/2$, and with reflection symmetry with respect to the $|c|$ axis
($|c|$ in the diagram is even in $F$). 

\acknowledgments
This work has been supported by the Spanish MCyT through grants BFM2002-04127-C02, by
the Third Regional Research Program of the Autonomous Region of Madrid (Strategic
Groups Action), and by the European Union under grant HPRN-CT-2002-00282. 

\appendix
\section{Nonlinear model having a explicit wavy wave front profile}
It is fairly easy to find smooth $g(u)$ having wave front solutions
with undamped oscillatory tails provided their dynamics is conservative. Let us
choose a profile of the form:
\begin{eqnarray} 
w(x)=\left\{ \begin{array}{ll} -1+ k_1 e^{ax} & x \leq 0 \\ 1+ k_2
\cos(bx+c) & x \geq 0 \end{array} \right. \nonumber
\end{eqnarray} 
This profile is continuous if $-1+k_1=1+k_2\cos c$, differentiable if  $k_1
a=-k_2 b \sin d$ and twice differentiable if $k_1 a^2 = -k_2 b^2 \cos c$. We set
$c=\tan^{-1} (b/a)$, $k_1=2/(1+a^2/b^2)$ and $k_2=-k_1 a /(b\sin c)$. The only
restriction on $a$ and $b$ is $b+ \tan^{-1}(b/a)<\pi/2$, $a,b>0$ to ensure that
$\cos(bx+c)$ is monotone in $0<x<1$. The profile $w(x)$ satisfies:
\begin{eqnarray} 
w_{xx}= \left\{ \begin{array}{ll} a^2 (w+1)\quad \mbox{for} & w< k_1-1, \\ 
-b^2(w+1)\quad \mbox{for} &  w > k_1-1,  \end{array} \right.\nonumber \\
w(x+1)-2w(x)+w(x-1)= f(w(x)),	\nonumber
\end{eqnarray} 
with:
\begin{eqnarray} 
f(w)=\left\{\begin{array}{ll} 2(w+1)(\cosh a-1) & w< w_1\\
(w+1)e^{-a}+Y-2w & (w_1,w_2) \\
k_2\cos(W+b)+k_1 e^{{a\over b}(W-b-c)}-2w & (w_2,w_3) \\ 
2(w-1)(\cos b-1) &  w>w_3
\end{array}\right.\nonumber
\end{eqnarray} 
Here, $W= k_2^{-1}\cos^{-1}(w-1) $, $Y=k_2\cos\{b a^{-1} \ln[(w+1)/k_1]+b+c\}$, $w_1=
e^{-a} k_1 -1$, $w_2= k_1 -1$, and $w_3= k_2 \cos(b+c) -1$. The continuous function
$f(w)$ is convex but, for a certain interval of values of $c^2$, the function
$g_c(u)= c^2 u_{xx} - f(u)$ has three zeroes and is strongly asymmetric. For these
values of $c$, the system $d^2u_n/dt^2 = u_{n+1}-2u_n+u_{n-1}-g_c(u_n)$ has a
traveling wave solution $u_n(t)=w(n-ct)$.


\begin{thebibliography}{}
\bibitem{definition}
In the conservative limit $m\to\infty$, the oscillations at one of both tails of the
wave front do not decay as $n\to\pm\infty$. This means that a slightly more general
definition of wave front solution is needed in the conservative case: $u_n(\tau)= w
(n-c\tau)$ joins  the neighborhoods of $U_1(F/A)$ and $U_3(F/A)$ (or viceversa) as $n$
increases from $-\infty$ to $\infty$. An infinitesimal amount of damping causes the
oscillatory tails of the wave front to decay to either $U_1(F/A)$ or $U_3(F/A)$ as
$n\to\pm\infty$.
\bibitem{numerical}
We would like to point out that the numerical solution of Eq.\ (\ref{cab1}) should be
calculated by using a scheme of sufficiently high order and small tolerance for the
cases $\alpha=0$ or $\alpha>0$ small (typically a standard Rung-Kutta-Fehlberg scheme
of orders 4/5 with tolerance $10^{-5}$ would do). Schemes with lower order or larger
tolerances have been found to produce spurious traveling waves. For the purpose of
checking the appropriatedness of a given numerical scheme, the exact solutions we
have discussed are invaluable.
\end{thebibliography}
\end{document}